\documentstyle[12pt]{article}
\date{}
\author{V.Dryuma*,~L.Bogdanov** \\[5mm]
{*IMI AS RM,} \\ {\it 5 Academiei Street, 2028
Kishinev, Moldova},\\[3mm]
{**L.D. Landau ITP RAS,} \\
{\it Kosygin str. 2,
Moscow 119334, Russia}
 }
\title{ON NONLINEAR EQUATIONS CONNECTED \\[2mm] WITH SIX-DIMENSIONAL PLEBANSKI SPACE}

\begin{document}
\maketitle

\begin{abstract}
\ \ \ \ An examples of Monge-Ampere equations connected with
six-dimensional generalization of the Plebanski four-dimensional
space are considered.
\end{abstract}

\section{The four-dimensional second heavenly equation}

     Plebanski second heavenly equation is connected with the metric of the form
\begin{equation}\label{dryuma:eq1}
ds^2=C(x,y,z,u){{\it dz}}^{2}+2\,B(x, y,z,u){\it dz}\,{\it
du}+A(x,y,z,u){{\it du}}^{2}+{\it dx}\,{\it du}+{\it dy}\,{\it dz}
\end{equation}

   Conditions for the metric ~(\ref{dryuma:eq1}) to be Ricci-flat
   \[
   R_{ik}=0
   \]
   lead to the equations on the coefficients
   \[
  R_{xz}=R_{yz}= {\frac {\partial }{\partial x}}B(x,y,z,u)+{\frac {\partial }{\partial
y}}C(x,y,z,u)=0,
\]
\[R_{xu}=R_{yu}= {\frac {\partial }{\partial
x}}A(x,y,z,u)+{\frac {\partial }{\partial y}}B(x,y,z,u)=0.
\]
\[
R_{zz}=2\,\left ({\frac {\partial }{\partial x}}A(x,y,z,u)\right
){\frac {
\partial }{\partial x}}C(x,y,z,u)+2\,A(x,y,z,u){\frac {\partial ^{2}}{
\partial {x}^{2}}}C(x,y,z,u)+\]\[+2\,\left ({\frac {\partial }{\partial x}}
B(x,y,z,u)\right ){\frac {\partial }{\partial
y}}C(x,y,z,u)+4\,B(x,y,z ,u){\frac {\partial ^{2}}{\partial
x\partial y}}C(x,y,z,u)+2\,{\frac {
\partial ^{2}}{\partial x\partial z}}B(x,y,z,u)-\]\[-2\,{\frac {\partial ^{
2}}{\partial u\partial x}}C(x,y,z,u)-2\,\left ({\frac {\partial }{
\partial x}}B(x,y,z,u)\right )^{2}+2\,C(x,y,z,u){\frac {\partial ^{2}}
{\partial {y}^{2}}}C(x,y,z,u)-\]\[-2\,\left ({\frac {\partial
}{\partial x} }C(x,y,z,u)\right ){\frac {\partial }{\partial
y}}B(x,y,z,u)=0,
\]
\[
R_{zu}=2\,A(x,y,z,u){\frac {\partial ^{2}}{\partial
{x}^{2}}}B(x,y,z,u)+2\, \left ({\frac {\partial }{\partial
x}}B(x,y,z,u)\right ){\frac {
\partial }{\partial y}}B(x,y,z,u)+\]\[+4\,B(x,y,z,u){\frac {\partial ^{2}}{
\partial x\partial y}}B(x,y,z,u)+{\frac {\partial ^{2}}{\partial x
\partial z}}A(x,y,z,u)-{\frac {\partial ^{2}}{\partial u\partial x}}B(
x,y,z,u)-\]\[-2\,\left ({\frac {\partial }{\partial
x}}C(x,y,z,u)\right ){ \frac {\partial }{\partial
y}}A(x,y,z,u)+2\,C(x,y,z,u){\frac {
\partial ^{2}}{\partial {y}^{2}}}B(x,y,z,u)-\]\[-{\frac {\partial ^{2}}{
\partial y\partial z}}B(x,y,z,u)+{\frac {\partial ^{2}}{\partial u
\partial y}}C(x,y,z,u)=0,
\]
\[
R_{uu}=2\,A(x,y,z,u){\frac {\partial ^{2}}{\partial
{x}^{2}}}A(x,y,z,u)+4\,B( x,y,z,u){\frac {\partial ^{2}}{\partial
x\partial y}}A(x,y,z,u)+\]\[+2\, \left ({\frac {\partial
}{\partial y}}B(x,y,z,u)\right ){\frac {
\partial }{\partial x}}A(x,y,z,u)+2\,\left ({\frac {\partial }{
\partial y}}C(x,y,z,u)\right ){\frac {\partial }{\partial y}}A(x,y,z,u
)+\]\[+2\,C(x,y,z,u){\frac {\partial ^{2}}{\partial
{y}^{2}}}A(x,y,z,u)+2\, {\frac {\partial ^{2}}{\partial u\partial
y}}B(x,y,z,u)-2\,{\frac {
\partial ^{2}}{\partial y\partial z}}A(x,y,z,u)-\]\[-2\,\left ({\frac {
\partial }{\partial x}}B(x,y,z,u)\right ){\frac {\partial }{\partial y
}}A(x,y,z,u)-2\,\left ({\frac {\partial }{\partial y}}B(x,y,z,u)
\right )^{2}=0.
\]

    After the substitution
\[C(x,y,z,u)=-{\frac {\partial ^{2}}{\partial {x}^{2}}}\theta(x,y,z,u)
,\quad B(x,y,z,u)={\frac {\partial ^{2}}{\partial x\partial
y}}\theta(x,y,z,u ),
\]
\[
A(x,y,z,u)=-{\frac {\partial
^{2}}{\partial {y}^{2}}}\theta(x,y,z,u)
\]
metric ~(\ref{dryuma:eq1}) takes the form
\[
ds^2=-\left ({\frac {\partial ^{2}}{
\partial {x}^{2}}}\theta(x,y,z,u)\right ){{\it dz}}^{2}+2\,\left ({
\frac {\partial ^{2}}{\partial x\partial y}}\theta(x,y,z,u)\right
){ \it dz}\,{\it du}-\left ({\frac {\partial ^{2}}{\partial
{y}^{2}}} \theta(x,y,z,u)\right ){{\it du}}^{2}+{\it dx}\,{\it
du}+{\it dy}\,{\it dz}.
\]

    It is Ricci-flat if the function $\theta(x,y,z,u)$ satisfies
the second Plebanski equation
\begin{equation}\label{dryuma:eq2}
    {\frac {\partial ^{2}}{\partial u\partial x}}\theta(x,y,z,u)+{\frac {
\partial ^{2}}{\partial y\partial z}}\theta(x,y,z,u)+\left ({\frac {
\partial ^{2}}{\partial {y}^{2}}}\theta(x,y,z,u)\right ){\frac {
\partial ^{2}}{\partial {x}^{2}}}\theta(x,y,z,u)-\left ({\frac {
\partial ^{2}}{\partial x\partial y}}\theta(x,y,z,u)\right
)^{2}=0
 \end{equation}
 \cite{dryuma:dr2}.

\section{Six-dimensional generalization }

    We introduce the following six-dimensional generalization of the metric ~(\ref{dryuma:eq1}),
\begin{equation}\label{dryuma:eq3}
{{\it ds}}^{2}={\it dx}\,{\it du}+{\it dy}\,{\it dv}+{\it
dz}\,{\it dw }+A(x,y,z,u,v,w){{\it
du}}^{2}+\]\[+2\,B(x,y,z,u,v,w){\it du}\,{\it dv}+2\,
E(x,y,z,u,v,w){\it du}\,{\it dw}+C(x,y,z,u,v,w){{\it
dv}}^{2}+\]\[+2\,H(x,y ,z,u,v,w){\it dv}\,{\it
dw}+F(x,y,z,u,v,w){{\it dw}}^{2}.
\end{equation}

The Ricci tensor of the metric ~(\ref{dryuma:eq3}) has fifteen components.
Nine of them  are equal to zero due the conditions
\begin{equation}\label{dryuma:eq4}
   {\frac {\partial }{\partial u}}E(x,y,z,u,v,w)+{\frac {\partial }{
\partial v}}H(x,y,z,u,v,w)+{\frac {\partial }{\partial w}}F(x,y,z,u,v,
w)=0,\]\[{\frac {\partial }{\partial u}}B(x,y,z,u,v,w)+{\frac
{\partial }{
\partial v}}C(x,y,z,u,v,w)+{\frac {\partial }{\partial w}}H(x,y,z,u,v,
w)=0,\]\[{\frac {\partial }{\partial u}}A(x,y,z,u,v,w)+{\frac
{\partial }{
\partial v}}B(x,y,z,u,v,w)+{\frac {\partial }{\partial w}}E(x,y,z,u,v.
w)=0.
\end{equation}

   This system of equation has solutions  depending  on arbitrary functions.

   In  a simplest case we have the solution
\[
A(x,y,z,u,v,w)=\left ({\frac {\partial ^{2}}{\partial
{z}^{2}}}f(x,y,z ,u,v,w)\right ){\frac {\partial ^{2}}{\partial
{y}^{2}}}f(x,y,z,u,v,w) -\left ({\frac {\partial ^{2}}{\partial
y\partial z}}f(x,y,z,u,v,w) \right )^{2},\]\[ C(x,y,z,u,v,w)=\left
({\frac {\partial ^{2}}{\partial {x}^{2}}}f(x,y,z ,u,v,w)\right
){\frac {\partial ^{2}}{\partial {z}^{2}}}f(x,y,z,u,v,w) -\left
({\frac {\partial ^{2}}{\partial x\partial z}}f(x,y,z,u,v,w)
\right )^{2},\]\[ F(x,y,z,u,v,w)=\left ({\frac {\partial
^{2}}{\partial {x}^{2}}}f(x,y,z ,u,v,w)\right ){\frac {\partial
^{2}}{\partial {y}^{2}}}f(x,y,z,u,v,w) -\left ({\frac {\partial
^{2}}{\partial x\partial y}}f(x,y,z,u,v,w) \right )^{2},\]\[\quad
E(x,y,z,u,v,w)=\left ({\frac {\partial ^{2}}{\partial y\partial
z}}f(x ,y,z,u,v,w)\right ){\frac {\partial ^{2}}{\partial
x\partial y}}f(x,y, z,u,v,w)-\]\[-\left ({\frac {\partial
^{2}}{\partial x\partial z}}f(x,y,z,u ,v,w)\right ){\frac
{\partial ^{2}}{\partial {y}^{2}}}f(x,y,z,u,v,w)
\]\[B(x,y,z,u,v,w)=\left ({\frac {\partial ^{2}}{\partial
x\partial z}}f(x ,y,z,u,v,w)\right ){\frac {\partial
^{2}}{\partial y\partial z}}f(x,y, z,u,v,w)-\]\[-\left ({\frac
{\partial ^{2}}{\partial x\partial y}}f(x,y,z,u ,v,w)\right
){\frac {\partial ^{2}}{\partial {z}^{2}}}f(x,y,z,u,v,w)\]\[
H(x,y,z,u,v,w)=\left ({\frac {\partial ^{2}}{\partial x\partial
z}}f(x ,y,z,u,v,w)\right ){\frac {\partial ^{2}}{\partial
x\partial y}}f(x,y, z,u,v,w)-\]\[-\left ({\frac {\partial
^{2}}{\partial y\partial z}}f(x,y,z,u ,v,w)\right ){\frac
{\partial ^{2}}{\partial {x}^{2}}}f(x,y,z,u,v,w)
\]
depending on one arbitrary function.

      Corresponding  six-dimensional metric looks as
\begin{equation}\label{dryuma:eq5}
^{6}ds^2=\left ({\frac {\partial ^{2}}{\partial {w}^{2}}}K(\vec x)
{\frac {\partial ^{2}}{\partial {v}^{2}}}K(\vec x)-\left ({\frac
{\partial ^{2}}{\partial v\partial w}}K(\vec x)\right )^{ 2}\right
){d{{x}}}^{2}\!+\!\]\[+2\,\left ({\frac {\partial ^{2}}{
\partial u\partial w}}K(\vec x){\frac {\partial ^{2}}{
\partial v\partial w}}K(\vec x)\!-\!{\frac {\partial ^{2}}{
\partial {w}^{2}}}K(\vec x){\frac {\partial ^{2}}{
\partial u\partial v}}K(\vec x)\right )d{{x}}d{{y}}\!+\!\]\[\!+\!\left (
{\frac {\partial ^{2}}{\partial {u}^{2}}}K(\vec x) {\frac
{\partial ^{2}}{\partial {w}^{2}}}K(\vec x)\!-\!\left ({\frac
{\partial ^{2}}{\partial u\partial w}}K(\vec x)\right )^{2} \right
){d{{y}}}^{2}\!+\!\]\[+2\,\left ({\frac {\partial ^{2}}{
\partial v\partial w}}K(\vec x){\frac {\partial ^{2}}{
\partial u\partial v}}K(\vec x)\!-\!{\frac {\partial ^{2}}{
\partial u\partial w}}K(\vec x){\frac {\partial ^{2}}{
\partial {v}^{2}}}K(\vec x)\right )d{{x}}d{{z}}\!+\!\]\[\!+\!\left ({
\frac {\partial ^{2}}{\partial {v}^{2}}}K(\vec x){\frac {
\partial ^{2}}{\partial {u}^{2}}}K(\vec x)\!-\!\left ({\frac {
\partial ^{2}}{\partial u\partial v}}K(\vec x)\right )^{2}\right
){d{{z}}}^{2}\!+\!\]\[\!+\!2\,\left ({\frac {\partial
^{2}}{\partial u
\partial w}}K(\vec x){\frac {\partial ^{2}}{\partial u
\partial v}}K(\vec x)\!-\!{\frac {\partial ^{2}}{\partial {u}^
{2}}}K(\vec x){\frac {\partial ^{2}}{\partial v\partial w }}K(\vec
x)\right )d{{z}}d{{y}}\!+\!\]\[\!+\!d{{x}}d{{u}}+d{{y}}d{{v}}+d
{{z}}d{{w}}
\end{equation}
where $K(\vec x)=K(x,y,z,u,v,w)$ is an arbitrary function.

   The Ricci tensor $R_{ij}$ of the metric ~(\ref{dryuma:eq5}) has six components.

   All equations
   \[
   R_{ij}=0
   \]
   after the substitution
   \[
   K(x,y,z,u,v,w)=\phi(y+v+x,z+w+x)
\]
 are reduced to one equation
\begin{equation}\label{dryuma:eq6}
-\left ({\frac {\partial ^{4}}{\partial \xi\partial {\rho}^{2}
\partial \xi}}\phi(\xi,\rho)\right ){\frac {\partial ^{2}}{\partial {
\xi}^{2}}}\phi(\xi,\rho)+2\,\left ({\frac {\partial ^{3}}{\partial
\xi
\partial \rho\partial \xi}}\phi(\xi,\rho)\right )^{2}-2\,\left ({
\frac {\partial ^{3}}{\partial {\rho}^{2}\partial
\xi}}\phi(\xi,\rho) \right ){\frac {\partial ^{3}}{\partial
{\xi}^{3}}}\phi(\xi,\rho)-\]\[- \left ({\frac {\partial
^{2}}{\partial {\rho}^{2}}}\phi(\xi,\rho) \right ){\frac {\partial
^{4}}{\partial {\xi}^{4}}}\phi(\xi,\rho)+2\, \left ({\frac
{\partial ^{2}}{\partial \rho\partial \xi}}\phi(\xi,\rho )\right
){\frac {\partial ^{4}}{\partial {\xi}^{2}\partial \rho
\partial \xi}}\phi(\xi,\rho)+\]\[+2\,\left ({\frac {\partial ^{3}}{
\partial {\rho}^{2}\partial \xi}}\phi(\xi,\rho)\right )^{2}-\left ({
\frac {\partial ^{4}}{\partial {\rho}^{4}}}\phi(\xi,\rho)\right ){
\frac {\partial ^{2}}{\partial {\xi}^{2}}}\phi(\xi,\rho)-2\,\left
({ \frac {\partial ^{3}}{\partial {\rho}^{3}}}\phi(\xi,\rho)\right
){ \frac {\partial ^{3}}{\partial \xi\partial \rho\partial
\xi}}\phi(\xi, \rho)-\]\[-\left ({\frac {\partial ^{2}}{\partial
{\rho}^{2}}}\phi(\xi,\rho )\right ){\frac {\partial ^{4}}{\partial
\xi\partial {\rho}^{2}
\partial \xi}}\phi(\xi,\rho)+2\,\left ({\frac {\partial ^{2}}{
\partial \rho\partial \xi}}\phi(\xi,\rho)\right ){\frac {\partial ^{4}
}{\partial {\rho}^{3}\partial \xi}}\phi(\xi,\rho)=0,
\end{equation}
 where
\[
\xi=x+y+v,\quad \rho=z+x+w.
\]

    In compact form this equation can be rewritten as
    \[
    \Delta\psi(\xi,\rho)=0
    \]
where
\[
\psi=(\xi,~\rho)=\left ({\frac {\partial ^{2}}{\partial
{\xi}^{2}}}\phi(\xi,\rho) \right ){\frac {\partial ^{2}}{\partial
{\rho}^{2}}}\phi(\xi,\rho)- \left ({\frac {\partial ^{2}}{\partial
\xi\partial \rho}}\phi(\xi,\rho )\right )^{2}
\]
and
\[
\Delta=\frac{\partial^2}{ \partial \xi^2}+\frac{\partial^2}{
\partial \rho^2}
\]
is the Laplace operator.

     Its solutions give Ricci-flat examples of the metric
     ~(\ref{dryuma:eq5}).

\section{ The Beltrami parameters}

    Two invariant equations defined by the first
\[
\Delta \psi= g^{i j}\frac{\partial \psi}{ \partial
x^i}\frac{\partial \psi}{
\partial x^j}
\]
 and the second
\[
\Box \psi=g^{ij}\left(\frac{\partial^2}{ \partial x^i
\partial x^j}-\Gamma^k_{ij}\frac{\partial}{
\partial x^k}\right)\psi
\]
Beltrami parameters can be considered to investigate the properties of the metrics ~(\ref{dryuma:eq5}).

    For the metric ~(\ref{dryuma:eq5}) the equation $\Box \phi=0$ looks as
\begin{equation}\label{dryuma:eq7}
 {\frac {\partial ^{2}}{\partial u\partial x}}\phi(\vec x)+{
\frac {\partial ^{2}}{\partial w\partial z}}\phi(\vec x)+{ \frac
{\partial ^{2}}{\partial v\partial y}}\phi(\vec x)- \left ({\frac
{\partial ^{2}}{\partial {y}^{2}}}\phi(\vec x) \right )\left
({\frac {\partial ^{2}}{\partial {x}^{2}}}f(\vec x) \right ){\frac
{\partial ^{2}}{\partial {z}^{2}}}f(\vec x)+\]\[+2\, \left ({\frac
{\partial ^{2}}{\partial x\partial z}}\phi(\vec x) \right )\left
({\frac {\partial ^{2}}{\partial x\partial z}}f(\vec x)\right
){\frac {\partial ^{2}}{\partial {y}^{2}}}f(\vec x)-2 \,\left
({\frac {\partial ^{2}}{\partial x\partial z}}\phi(\vec x)\right
)\left ({\frac {\partial ^{2}}{\partial y\partial z}}f(\vec
x)\right ){\frac {\partial ^{2}}{\partial x\partial y}}f(\vec
x)+\]\[+2\,\left ({\frac {\partial ^{2}}{\partial x\partial
y}}\phi(\vec x)\right )\left ({\frac {\partial ^{2}}{\partial
x\partial y}}f(\vec x)\right ){\frac {\partial ^{2}}{\partial
{z}^{2}}}f(\vec x)-\left ({\frac {\partial ^{2}}{\partial
{x}^{2}}}\phi(\vec x )\right )\left ({\frac {\partial
^{2}}{\partial {z}^{2}}}f(\vec x )\right ){\frac {\partial
^{2}}{\partial {y}^{2}}}f(\vec x)+\]\[+2\, \left ({\frac {\partial
^{2}}{\partial y\partial z}}\phi(\vec x) \right )\left ({\frac
{\partial ^{2}}{\partial y\partial z}}f(\vec x)\right ){\frac
{\partial ^{2}}{\partial {x}^{2}}}f(\vec x)-2 \,\left ({\frac
{\partial ^{2}}{\partial y\partial z}}\phi(\vec x )\right )\left
({\frac {\partial ^{2}}{\partial x\partial z}}f(\vec x)\right
){\frac {\partial ^{2}}{\partial x\partial y}}f(\vec x)-\]\[-\left
({\frac {\partial ^{2}}{\partial {z}^{2}}}\phi(\vec x)\right
)\left ({\frac {\partial ^{2}}{\partial {x}^{2}}}f(\vec x)\right
){\frac {\partial ^{2}}{\partial {y}^{2}}}f(\vec x)-2\, \left
({\frac {\partial ^{2}}{\partial x\partial y}}\phi(\vec x) \right
)\left ({\frac {\partial ^{2}}{\partial x\partial z}}f(\vec
x)\right ){\frac {\partial ^{2}}{\partial y\partial z}}f(\vec
x)+\]\[+\left ({\frac {\partial ^{2}}{\partial {z}^{2}}}\phi(\vec
x) \right )\left ({\frac {\partial ^{2}}{\partial x\partial
y}}f(\vec x)\right )^{2}+\left ({\frac {\partial ^{2}}{\partial
{y}^{2}}} \phi(\vec x)\right )\left ({\frac {\partial
^{2}}{\partial x
\partial z}}f(\vec x)\right )^{2}+\left ({\frac {\partial ^{2}
}{\partial {x}^{2}}}\phi(\vec x)\right )\left ({\frac {\partial ^
{2}}{\partial y\partial z}}f(\vec x)\right )^{2}=0.
\end{equation}

   In a special  case  equation ~(\ref{dryuma:eq7}) after the substitution
   \[
   \phi(\vec x)=f(\vec x)
   \]
takes the form
\begin{equation}\label{dryuma:eq8}
{\frac {\partial ^{2}}{\partial u\partial x}}f(\vec x)+{ \frac
{\partial ^{2}}{\partial w\partial z}}f(\vec x)+{\frac {
\partial ^{2}}{\partial v\partial y}}f(\vec x)-3\,\left ({\frac
{\partial ^{2}}{\partial {y}^{2}}}f(\vec x)\right )\left ({\frac
{\partial ^{2}}{\partial {x}^{2}}}f(\vec x)\right ){\frac {
\partial ^{2}}{\partial {z}^{2}}}f(\vec x)+\]\[+3\,\left ({\frac {
\partial ^{2}}{\partial x\partial z}}f(\vec x)\right )^{2}{\frac
{\partial ^{2}}{\partial {y}^{2}}}f(\vec x)-6\,\left ({\frac {
\partial ^{2}}{\partial x\partial z}}f(\vec x)\right )\left ({
\frac {\partial ^{2}}{\partial y\partial z}}f(\vec x)\right ){
\frac {\partial ^{2}}{\partial x\partial y}}f(\vec
x)+\]\[+3\,\left ( {\frac {\partial ^{2}}{\partial x\partial
y}}f(\vec x)\right )^{2 }{\frac {\partial ^{2}}{\partial
{z}^{2}}}f(\vec x)+3\,\left ({ \frac {\partial ^{2}}{\partial
y\partial z}}f(\vec x)\right )^{2} {\frac {\partial ^{2}}{\partial
{x}^{2}}}f(\vec x)=0. \end{equation}

   After the change of variables
\[
f(\vec x)=f(x,y,z,u,v,w)=h(x+u,v+y,w+z)=h(\eta,\xi,\rho)
\]
equation ~(\ref{dryuma:eq8}) is reduced to the form
\begin{equation}\label{dryuma:eq9}
 {\frac {\partial
^{2}}{\partial {\eta}^{2}}}h(\eta,\xi,\rho)+{\frac {
\partial ^{2}}{\partial {\rho}^{2}}}h(\eta,\xi,\rho)+{\frac {\partial
^{2}}{\partial {\xi}^{2}}}h(\eta,\xi,\rho)+3\,\left ({\frac
{\partial ^{2}}{\partial \eta\partial \rho}}h(\eta,\xi,\rho)\right
)^{2}{\frac {
\partial ^{2}}{\partial {\xi}^{2}}}h(\eta,\xi,\rho)-\]\[-6\,\left ({\frac {
\partial ^{2}}{\partial \eta\partial \rho}}h(\eta,\xi,\rho)\right )
\left ({\frac {\partial ^{2}}{\partial \rho\partial
\xi}}h(\eta,\xi, \rho)\right ){\frac {\partial ^{2}}{\partial
\eta\partial \xi}}h(\eta, \xi,\rho)+3\,\left ({\frac {\partial
^{2}}{\partial \eta\partial \xi}} h(\eta,\xi,\rho)\right
)^{2}{\frac {\partial ^{2}}{\partial {\rho}^{2}
}}h(\eta,\xi,\rho)-\]\[-3\,\left ({\frac {\partial ^{2}}{\partial
{\xi}^{2} }}h(\eta,\xi,\rho)\right )\left ({\frac {\partial
^{2}}{\partial {\eta }^{2}}}h(\eta,\xi,\rho)\right ){\frac
{\partial ^{2}}{\partial {\rho}^ {2}}}h(\eta,\xi,\rho)+3\,\left
({\frac {\partial ^{2}}{\partial \rho
\partial \xi}}h(\eta,\xi,\rho)\right )^{2}{\frac {\partial ^{2}}{
\partial {\eta}^{2}}}h(\eta,\xi,\rho)=0
 \end{equation}
or
\[
\Delta h(\eta,\xi,\rho)-3\det\left [\begin {array}{ccc} {\frac
{\partial ^{2}}{
\partial {\eta}^{2}}}h(\eta,\xi,\rho)&{\frac {\partial ^{2}}{
\partial \eta\partial \xi}}h(\eta,\xi,\rho)&{\frac {\partial ^{2}}
{\partial \eta\partial \rho}}h(\eta,\xi,\rho)\\\noalign{\medskip}{
\frac {\partial ^{2}}{\partial \eta\partial \xi}}h(\eta,\xi,\rho)&
{\frac {\partial ^{2}}{\partial {\xi}^{2}}}h(\eta,\xi,\rho)&{
\frac {\partial ^{2}}{\partial \rho\partial \xi}}h(\eta,\xi,\rho)
\\\noalign{\medskip}{\frac {\partial ^{2}}{\partial \eta\partial
\rho}}h(\eta,\xi,\rho)&{\frac {\partial ^{2}}{\partial \rho
\partial \xi}}h(\eta,\xi,\rho)&{\frac {\partial ^{2}}{\partial {
\rho}^{2}}}h(\eta,\xi,\rho)\end {array}\right ]=0,
\]
where
\[ \Delta=\frac{\partial^2}{ \partial
\eta^2}+\frac{\partial^2}{ \partial \xi^2}+\frac{\partial^2}{
\partial \rho^2}
\]
is a three-dimensional Laplace operator.

    Solutions of equation ~(\ref{dryuma:eq9}) are characterized the properties of the metric (\ref{dryuma:eq5}).

\section{Particular solutions of equation  ~(\ref{dryuma:eq6})}

      To obtain particular solutions of  partial nonlinear
       differential equation
\begin{equation}\label{Dr3}
F(x,y,z,f_x,f_y,f_z,f_{xx},f_{xy},f_{xz},f_{yy},f_{yz},f_{xxx},f_{xyy},f_{xxy},..)=0,
 \end{equation}
    a following approach can be applied.

      We use parametric presentation of the functions and variables
\begin{equation}\label{Dr4}
f(x,y,z,s)\rightarrow u(x,t,z,s),\quad y \rightarrow
v(x,t,z,s),\quad f_x\rightarrow
u_x-\frac{u_t}{v_t}v_x,f_s\rightarrow \quad
u_s-\frac{u_t}{v_t}v_s,\]\[ f_z\rightarrow
u_z-\frac{u_t}{v_t}v_z,\quad f_y \rightarrow \frac{u_t}{v_t},
\quad f_{yy} \rightarrow \frac{(\frac{u_t}{v_t})_t}{v_t}, \quad
f_{xy} \rightarrow \frac{(u_x-\frac{u_t}{v_t}v_x)_t}{v_t},...
\end{equation}
where variable $t$ is considered as parameter.

  Note that conditions of the type
   \[
   f_{xy}=f_{yx},\quad f_{xz}=f_{zx},\quad f_{xs}=f_{sx}...
   \]
are fulfilled at the such type of presentation.

  As a result instead of equation (\ref{Dr3}) one gets the
  relation between new variables $u(x,t,z)$ and $v(x,t,z)$ and
  their partial derivatives
\begin{equation}\label{Dr5}
\Psi(u,v,u_x,u_z,u_t,u_s,v_x,v_z,v_t,v_s...)=0.
  \end{equation}

    This relation coincides with initial p.d.e. for $v(x,t,z,s)=t$
    and takes more general form after presentation of the functions $u,v$ in the form
    $u(x,t,z,,s)=F(\omega,\omega_t...)$,
    $v(x,t,z,s)=\Phi(\omega,\omega_t...)$ with some function $\omega(x,t,z,s)$  .

\subsection{The example. Laplace equation}

    Two-dimensional Laplace equation
    \begin{equation}\label{Dr6}
  {\frac {\partial ^{2}}{\partial {x}^{2}}}f(x,y)+{\frac {\partial ^{2}}
{\partial {y}^{2}}}f(x,y) =0 \end{equation}
 after
$(u,v)$-transformation with the conditions
\[
u(x,t)=t{\frac {\partial }{\partial
t}}\omega(x,t)-\omega(x,t),\quad
 v(x,t)={\frac {\partial
}{\partial t}}\omega(x,t)
\]
takes the form of Monge-Ampere equation
\begin{equation}\label{Dr7}
-\left ({\frac {\partial ^{2}}{\partial {t}^{2}}}\omega(x,t)\right
){ \frac {\partial ^{2}}{\partial {x}^{2}}}\omega(x,t)+\left
({\frac {
\partial ^{2}}{\partial t\partial x}}\omega(x,t)\right )^{2}+1=0.
\end{equation}

    Particular solution of this equation is
    \[
    \omega(x,t)=A(t)+x+{x}^{2}C(t)
\]
where
\[
A(t)=-1/12\,{\it \_C1}\,{t}^{3}+{\it \_C4}
\]
and
\[
C(t)=-{\frac {1}{{\it \_C1}\,t}}.
\]

     The elimination of the parameter $t$ from the relations
   \[4\,y{\it \_C1}\,{t}^{2}+{{\it \_C1}}^{2}{t}^{4}-4\,{x}^{2}
=0,
\]
\[\quad 6\,f(x,y){\it \_C1}\,t+{{\it
\_C1}}^{2}{t}^{4}-12\,{x}^{2}+6\,{\it \_C4}\,{\it \_C1}\,t+6\,{\it
\_C1}\,tx =0
\]
gives us the function
\[f(x,y)=\]\[=1/18\,{\frac {-18\,{\it \_C4}\,{\it \_C1}-18\,x{\it \_C1}+12\,
\sqrt {-2\,{y}^{3}{\it \_C1}+2\,{\it \_C1}\,\sqrt
{{x}^{6}+3\,{x}^{4}{
y}^{2}+3\,{y}^{4}{x}^{2}+{y}^{6}}+6\,{x}^{2}y{\it \_C1}}}{{\it
\_C1}}}
\]
 satisfying  two-dimensional Laplace equation ~(\ref{Dr6}).

More general solutions of  equation ~(\ref{Dr6}) can be also constructed
    from solutions of equation ~(\ref{Dr7}) in a similar way.

    Note that such type of solutions of  Laplace equation
    may be applied in the theory of water waves.

\medskip

    To construct particular solutions of equation ~(\ref{dryuma:eq6})
\[
-\left ({\frac {\partial ^{4}}{\partial y\partial {x}^{2}\partial
y}} \phi(x,y)\right ){\frac {\partial ^{2}}{\partial
{x}^{2}}}\phi(x,y)+2 \,\left ({\frac {\partial ^{3}}{\partial
{x}^{2}\partial y}}\phi(x,y) \right )^{2}-2\,\left ({\frac
{\partial ^{3}}{\partial y\partial x
\partial y}}\phi(x,y)\right ){\frac {\partial ^{3}}{\partial {x}^{3}}}
\phi(x,y)-\]\[-\left ({\frac {\partial ^{2}}{\partial
{y}^{2}}}\phi(x,y) \right ){\frac {\partial ^{4}}{\partial
{x}^{4}}}\phi(x,y)+2\,\left ({ \frac {\partial ^{2}}{\partial
x\partial y}}\phi(x,y)\right ){\frac {
\partial ^{4}}{\partial {x}^{3}\partial y}}\phi(x,y)+\]\[+2\,\left ({\frac
{\partial ^{3}}{\partial y\partial x\partial y}}\phi(x,y)\right
)^{2}- \left ({\frac {\partial ^{4}}{\partial
{y}^{4}}}\phi(x,y)\right ){ \frac {\partial ^{2}}{\partial
{x}^{2}}}\phi(x,y)-2\,\left ({\frac {
\partial ^{3}}{\partial {y}^{3}}}\phi(x,y)\right ){\frac {\partial ^{3
}}{\partial {x}^{2}\partial y}}\phi(x,y)-\]\[-\left ({\frac
{\partial ^{2}} {\partial {y}^{2}}}\phi(x,y)\right ){\frac
{\partial ^{4}}{\partial y
\partial {x}^{2}\partial y}}\phi(x,y)+2\,\left ({\frac {\partial ^{2}}
{\partial x\partial y}}\phi(x,y)\right ){\frac {\partial ^{4}}{
\partial {y}^{2}\partial x\partial y}}\phi(x,y)
=0.
\]
we use the method described above.

    After the transformation of the function $\phi(x,y)$ and its derivatives
in accordance with the rules (\ref{Dr4}) and substitution of
corresponding expressions into equation (\ref{dryuma:eq6}) one
obtains relation of the type (\ref{Dr5}).

    From this relation  in a simplest case
     \[
     u(x,t)=A(t)x+{x}^{2},\quad
v(x,t)=B(t)x+x ,\quad B(t)=t
\]
 we get the equation for the function $A(t)$
\[
\left ({\frac {d^{4}}{d{t}^{4}}}A(t)\right ){t}^{2}+2\,\left
({\frac { d^{4}}{d{t}^{4}}}A(t)\right )t+2\,{\frac
{d^{4}}{d{t}^{4}}}A(t)+4\, \left ({\frac
{d^{3}}{d{t}^{3}}}A(t)\right )t+4\,{\frac {d^{3}}{d{t}^{
3}}}A(t)+2\,{\frac {d^{2}}{d{t}^{2}}}A(t)=0,
\]
having a general solution
\begin{equation}\label{Dr8}
A(t)={\it \_C1}+{\it \_C2}\,t+{\it \_C3}\,\left
(\arctan(t+1)t+\arctan (t+1)-1/2\,\ln ({t}^{2}+2\,t+2)\right
)+\]\[+{\it \_C4}\,\left (1/2\,t\ln ( {t}^{2}+2\,t+2)+\ln
({t}^{2}+2\,t+2)-\arctan(t+1)t\right ).
\end{equation}

     Elimination of the parameter $t$ from the conditions
      \[
      \phi(x,y)-(A(t)x+x^2)=0,\quad
      y-tx-x=0
      \]
with the function $A(t)$ from ~(\ref{Dr8}) leads to the solution of
equation (\ref{dryuma:eq6})
\[
\phi(x,y)=x{\it \_C1}+{\it \_C2}\,y-{\it \_C2}\,x+{\it
\_C3}\,\arctan({\frac {y} {x}})y-1/2\,x{\it \_C3}\,\ln ({\frac
{{y}^{2}+{x}^{2}}{{x}^{2}}})+\]\[+1/2 \,{\it \_C4}\,\ln ({\frac
{{y}^{2}+{x}^{2}}{{x}^{2}}})y+1/2\,x{\it \_C4}\,\ln ({\frac
{{y}^{2}+{x}^{2}}{{x}^{2}}})-{\it \_C4}\,\arctan({ \frac
{y}{x}})y+{\it \_C4}\,\arctan({\frac {y}{x}})x+{x}^{2}.
\]

\section{Particular solutions of equation
~(\ref{dryuma:eq9})}

    Equation
    ~(\ref{dryuma:eq9}), rewritten in new notations
\begin{equation}\label{Dr8a}
{\frac {\partial ^{2}}{\partial {x}^{2}}}h(x,y,z)+{\frac {\partial
^{2 }}{\partial {z}^{2}}}h(x,y,z)+{\frac {\partial ^{2}}{\partial
{y}^{2}} }h(x,y,z)+3\,\left ({\frac {\partial ^{2}}{\partial
x\partial z}}h(x,y ,z)\right )^{2}{\frac {\partial ^{2}}{\partial
{y}^{2}}}h(x,y,z)-\]\[-6\, \left ({\frac {\partial ^{2}}{\partial
x\partial z}}h(x,y,z)\right ) \left ({\frac {\partial
^{2}}{\partial y\partial z}}h(x,y,z)\right ){ \frac {\partial
^{2}}{\partial x\partial y}}h(x,y,z)+3\,\left ({\frac {\partial
^{2}}{\partial x\partial y}}h(x,y,z)\right )^{2}{\frac {
\partial ^{2}}{\partial {z}^{2}}}h(x,y,z)-\]\[-3\,\left ({\frac {\partial ^
{2}}{\partial {y}^{2}}}h(x,y,z)\right )\left ({\frac {\partial
^{2}}{
\partial {x}^{2}}}h(x,y,z)\right ){\frac {\partial ^{2}}{\partial {z}^
{2}}}h(x,y,z)+3\,\left ({\frac {\partial ^{2}}{\partial y\partial
z}}h (x,y,z)\right )^{2}{\frac {\partial ^{2}}{\partial
{x}^{2}}}h(x,y,z)=0, \end{equation}
 can be transformed   into
the form
\begin{equation}\label{Dr9}
-3\,\left ({\frac {\partial ^{2}}{\partial {x}^{2}}}\omega(x,t,z)
\right ){\frac {\partial ^{2}}{\partial {z}^{2}}}\omega(x,t,z)+3\,
\left ({\frac {\partial ^{2}}{\partial x\partial z}}\omega(x,t,z)
\right )^{2}-\left ({\frac {\partial ^{2}}{\partial
{t}^{2}}}\omega(x, t,z)\right ){\frac {\partial ^{2}}{\partial
{x}^{2}}}\omega(x,t,z)+\]\[+1+ \left ({\frac {\partial
^{2}}{\partial t\partial z}}\omega(x,t,z) \right )^{2}+\left
({\frac {\partial ^{2}}{\partial t\partial x}} \omega(x,t,z)\right
)^{2}-\left ({\frac {\partial ^{2}}{\partial {t}^{
2}}}\omega(x,t,z)\right ){\frac {\partial ^{2}}{\partial {z}^{2}}}
\omega(x,t,z)=0,
\end{equation}
 according to the rules (\ref{Dr4})
and using the substitution
\[
u(x,t,z)=t{\frac {\partial }{\partial
t}}\omega(x,t,z)-\omega(x,t,z),\quad
 v(x,t,z)={\frac {\partial
}{\partial t}}\omega(x,t,z).
\]

    In particular case
\[
\omega(x,t,z)=A({x}^{2}+{z}^{2},t)
\]
equation (\ref{Dr9}) takes the form
\begin{equation}\label{Dr10}
-24\,\left ({\frac {\partial ^{2}}{\partial
{\xi}^{2}}}A(\xi,t)\right )\left ({\frac {\partial }{\partial
\xi}}A(\xi,t)\right )\xi-12\, \left ({\frac {\partial }{\partial
\xi}}A(\xi,t)\right )^{2}-4\,\left ({\frac {\partial
^{2}}{\partial {t}^{2}}}A(\xi,t)\right )\left ({ \frac {\partial
^{2}}{\partial {\xi}^{2}}}A(\xi,t)\right )\xi-\]\[-4\, \left
({\frac {\partial ^{2}}{\partial {t}^{2}}}A(\xi,t)\right ){ \frac
{\partial }{\partial \xi}}A(\xi,t)+1+4\,\left ({\frac {\partial
^{2}}{\partial t\partial \xi}}A(\xi,t)\right )^{2}\xi=0,
\end{equation}
 where $\xi=x^2+z^2$.

    This type of equations meet in the theory of turbulent flow ~\cite{rubts}.

    Particular solution of equation (\ref{Dr10}) is of the form
    \[
    A(\xi,t)=B(t)+\xi\,{e^{Ct}},
\]
where the function $B(t)$ is defined by the expression
\[
B(t)=1/4\,{\frac {{e^{-Ct}}}{{C}^{2}}}-3\,{\frac
{{e^{Ct}}}{{C}^{2}}}+ {\it \_C1}\,t+{\it \_C2}.
\]
      Now elimination of the parameter $t$ from the system of equations
      \[
h(x,y,z)-t{\frac {\partial }{\partial
t}}\omega(x,t,z)+\omega(x,t,z)=0,
\]
\[ y-{\frac {\partial
}{\partial t}}\omega(x,t,z)=0.
\]
leads in the case $\_C1=0,~\_C2=0,~C=1$ to the function
\[
h(x,y,z)=\left (y\sqrt {{y}^{2}-3+{x}^{2}+{z}^{2}}+{y}^{2}\right
)\ln ({\frac {y+\sqrt
{{y}^{2}-3+{x}^{2}+{z}^{2}}}{-3+{x}^{2}+{z}^{2}}}) \left (y+\sqrt
{{y}^{2}-3+{x}^{2}+{z}^{2}}\right )^{-1}+\]\[+{\frac {3-{x}^
{2}-{y}^{2}-{z}^{2}-\ln (2)y\sqrt {{y}^{2}-3+{x}^{2}+{z}^{2}}-\ln
(2){ y}^{2}-y\sqrt {{y}^{2}-3+{x}^{2}+{z}^{2}}}{y+\sqrt
{{y}^{2}-3+{x}^{2}+ {z}^{2}}}}
\]
satisfying the equation (\ref{Dr8a}).

    Another type of particular solutions of equation ~(\ref{Dr8a})
    can be obtained using the reduction  of equation ~(\ref{Dr9}) after the substitution
    \[
    \omega(x,t,z)=A(x+t,z)
\]
to the form
   \begin{equation}\label{Dr11}
   \left ({\frac {\partial ^{2}}{\partial {x}^{2}}}A(x,y)\right ){\frac {
\partial ^{2}}{\partial {y}^{2}}}A(x,y)-\left ({\frac {\partial ^{2}}{
\partial x\partial y}}A(x,y)\right )^{2}-1/4=0,
 \end{equation}
 where $A(x,y)=A(x+t,z)$.

      The $(u,v)$-transformation of equation ~(\ref{Dr11}) with the
      condition
\[
u(x,t)=t{\frac {\partial }{\partial
t}}\omega(x,t)-\omega(x,t),\quad
 v(x,t)={\frac {\partial
}{\partial t}}\omega(x,t)
\]
leads to Laplace equation for the function $\omega(x,t)$
\[
4\,{\frac {\partial ^{2}}{\partial {x}^{2}}}\omega(x,t)+{\frac {
\partial ^{2}}{\partial {t}^{2}}}\omega(x,t)=0.
\]

      Its general solution has the form
      \[
      \omega(x,t)=M(x+2\,I t)+N(x-2\,I t),
\]
where $M$ and $N$ are arbitrary functions.

     After the choice of the function $\omega(x,t)$ and elimination of the parameter $t$ from equations
\[
A(x,y)-(t{\frac {\partial }{\partial
t}}\omega(x,t)-\omega(x,t)=0,\quad
 y-{\frac {\partial
}{\partial t}}\omega(x,t)=0,
\]
the function $A(x,y)$ and then the function
    \[
    \omega(x,t,z)=A(x+t,z)
    \]
can be found.

     Elimination of parameter $t$ from equations
   \begin{equation}\label{Dr12}
h(x,y,z)-t{\frac {\partial }{\partial
t}}\omega(x,t,z)+\omega(x,t,z)=0,
\]
\[ y-{\frac {\partial
}{\partial t}}\omega(x,t,z)=0
\end{equation}
 with a given
function $\omega(x,t,z)$ allow us to obtain particular solution of
equation ~(\ref{Dr8a}).

    Let us consider an example.

    We take  solution of the Laplace equation  of the form
    \[
    \omega(x,t)=\left (x+2\,I t\right )^{2}+\left (x-2\,I t\right )^{3}
\]
or
\[
\omega(x,t)={x}^{2}-4\,{t}^{2}+{x}^{3}-12\,x{t}^{2}+I\left
(4\,tx-6\,t{x}^{2}+8\,{ t}^{3}\right ).
\]

     Its imaginary part
     \[
\omega_1= 4\,tx-6\,t{x}^{2}+8\,{t}^{3}
\]
 satisfies Laplace equation.

     From the conditions
     \[
     A(x,y)-t{\frac {\partial }{\partial
     t}}\omega_1(x,t)+\omega_1(x,t)=0,
\]
\[
y-{\frac {\partial }{\partial t}}\omega_1(x,t)=0
\]
we get the system of equations
\[
A(x,y)-16\,{t}^{3}=0,\quad y-4\,x+6\,{x}^{2}-24\,{t}^{2}=0
\]
from which we eliminate the parameter $t$.

    As a result we find the function $A(x,y)$
    \[
    A(x,y)=1/18\,\sqrt {6\,y-24\,x+36\,{x}^{2}}y-2/9\,\sqrt {6\,y-24\,x+36
\,{x}^{2}}x+1/3\,\sqrt {6\,y-24\,x+36\,{x}^{2}}{x}^{2}
\]
satisfying equation ~(\ref{Dr11}).

      From this function after the change of variables
    \[  x=x+t,\quad
y=z\]
 we find the function
\[
\omega(x,t,z)=1/18\,\sqrt
{6\,z-24\,x-24\,t+36\,{x}^{2}+72\,tx+36\,{t} ^{2}}\left
(z-4\,x-4\,t+6\,{x}^{2}+12\,tx+6\,{t}^{2}\right )
\]

     Using this function we eliminate the parameter $t$
     from relations ~(\ref{Dr12}) and obtain corresponding solution
\[
h(x,y,z)=-{\frac {1}{54}}\,{\frac {\left (-2-3\,\sqrt
{2-3\,z+T}+3\,z+ 9\,x\sqrt {2-3\,z+T}\right )T}{\sqrt
{-2+3\,z+T}}}-\]\[-{\frac {1}{54}}\,{ \frac
{\!-\!12\,z\!+\!6\,\sqrt {2\!-\!3\,z+T}\!-\!18\,x\sqrt
{2-3\,z\!+\!T}\!+\!4-\!\!9\,z\sqrt {2
-3\,z\!+\!T}\!+\!9\,{z}^{2}\!+\!27\,zx\sqrt
{2-3\,z\!+\!T}\!-\!36\,{y}^{2}}{\sqrt {\!-\!2+\!3\,z\!+\!T}}},
\]
where
\[
T=\sqrt {9\,{z}^{2}-12\,z+4+36\,{y}^{2}}.
\]

\section{Acknowledgements}

     The research was partially supported in the framework of joint Russian-Moldavian
research project
(Grant 08.820.08.07 RF of HCSTD ASM, Moldova, and RFBR grant 08-01-90104, Russia).

\end{document}